\documentclass[aps,prb,twocolumn,floatfix,amssymb]{revtex4}

\usepackage{graphicx}

\begin{document}

\title{Bipolar spin filter in a quantum dot molecule}

\author{F. Mireles}
\affiliation{Departamento de F{\'\i}sica Te\'orica, Centro de
Ciencias de la Materia Condensada -- Universidad Nacional
Aut\'onoma de M\'exico, Ensenada, Baja California, M\'exico
22800}

\author{S. E. Ulloa}
\affiliation{Department of Physics and Astronomy and Nanoscale and
Quantum Phenomena Institute, Ohio University, Athens, OH
45701-2979}

\author{E. Cota}
\affiliation{Departamento de F{\'\i}sica Te\'orica, Centro de
Ciencias de la Materia Condensada -- Universidad Nacional
Aut\'onoma de M\'exico, Ensenada, Baja California, M\'exico
22800}

\author{F. Rojas}
\affiliation{Departamento de F{\'\i}sica Te\'orica, Centro de
Ciencias de la Materia Condensada -- Universidad Nacional
Aut\'onoma de M\'exico, Ensenada, Baja California, M\'exico
22800}

\begin{abstract}
\vspace{0.2 in}
We show that the tunable hybridization between two lateral quantum
dots connected to non-magnetic current leads in a `hanging-dot'
configuration that can be used to implement a bipolar spin filter. The
competition between Zeeman, exchange interaction, and interdot tunneling
(molecular hybridization) yields a singlet-triplet transition of
the double dot {\it ground state} that allows spin filtering in
Coulomb blockade experiments. Its generic nature should make it
broadly useful as a robust bidirectional spin polarizer.
\end{abstract}

\pacs{73.63.Kv, 72.25.-b, 73.21.La, 72.25.Dc}

\maketitle

Controlling the spin of electrons in mesoscopic systems is an
important task in `spintronic devices,' as well as in the
fundamental understanding of spin relaxation and
coherence.\cite{Wolf} Experiments using diluted magnetic
semiconductors (DMS) heterostructures have reported excellent
polarization values ($\sim 90\%$). \cite{OhnoDMS} Spin-polarized
currents have been observed using ferromagnetic leads in
semiconductor quantum dots. \cite{ciorga} Pioneering studies have
also appeared recently where different spin-filter and
polarization measurements have been described where no
polarization of the leads is needed. The original proposal of
Recher {\em et al}. \cite{Recher} -- to produce a spin-polarized
current using the Zeeman effect in a quantum dot with odd number
of electrons -- has been recently implemented in beautiful
experiments by Hanson {\em et al}. \cite{Hanson} They report
nearly pure ($\sim 99\%$) spin collection for spin-unpolarized
leads, a reflection of the large orbital energy separation
achieved in their small quantum dots. Moreover, they can swiftly
flip the spin polarization (a `bipolar' filter) by changing the
charge state of the single dot. Bidirectional spin filtering
making clever use of spin coherence \cite{Folk-Science} and
magnetic focusing was reported by Potok {\em et al}.,
\cite{Potok1,Potok2} who achieved very good spin ($\sim 70\%$)
polarization at moderate applied fields ($\sim 6T$ parallel to the
plane of the dot).

Spin-blockade \cite{Ono} in double quantum dots (DQD)
connected in series (sequentially) and coupled via tunneling  has
also been studied by Johnson {\em et al}. \cite{Johnson} using
transport measurements and charge sensing with quantum point
contacts.  They observe current rectification due to the
singlet-triplet spin-blockade mechanism in the DQD.\@ Coulomb- and
spin-blockade spectroscopy studies in spin sensitive experiments
were also realized recently in a two-level DQD molecule.
\cite{Pioro}

In this letter we show theoretically that a bipolar spin filter
can be implemented at moderate fields by producing a
singlet-triplet transition of the DQD {\em ground state}.  This
transition is obtained solely by tuning the hybridization
(hopping) between the two quantum dots in the molecule, achieved
by varying the coupling via the quantum point contact connecting
the dots. The DQD is connected to non-magnetic leads via only one
of the dots, in a `hanging dot' configuration (Fig.\ 1a). It is
shown that the competition between Zeeman energy and the effective
`superexchange' interaction results in lower energy for the {\em
singlet} configuration for large enough interdot tunneling and
even electron number.  This in turns gives rise to a natural spin
selectivity, fully tunable by appropriate electrical gating of the
structure. The effect is shown to arise in both the linear (low
bias) and non-linear regimes of transport at low temperatures.

% fig 1

We model the coupled DQD molecule as a single coherent system,
where the quantum dots are assumed small enough so as to have only
one relevant orbital energy level on each dot with gate-controlled
on-site energies {$\epsilon_i$, ($i=1,2$). The conducting dot in
the molecule (dot 2 in Fig.\ 1a) is weakly coupled to non-magnetic
reservoirs.

\begin{figure}[htb]
\includegraphics[width=2.6 in]{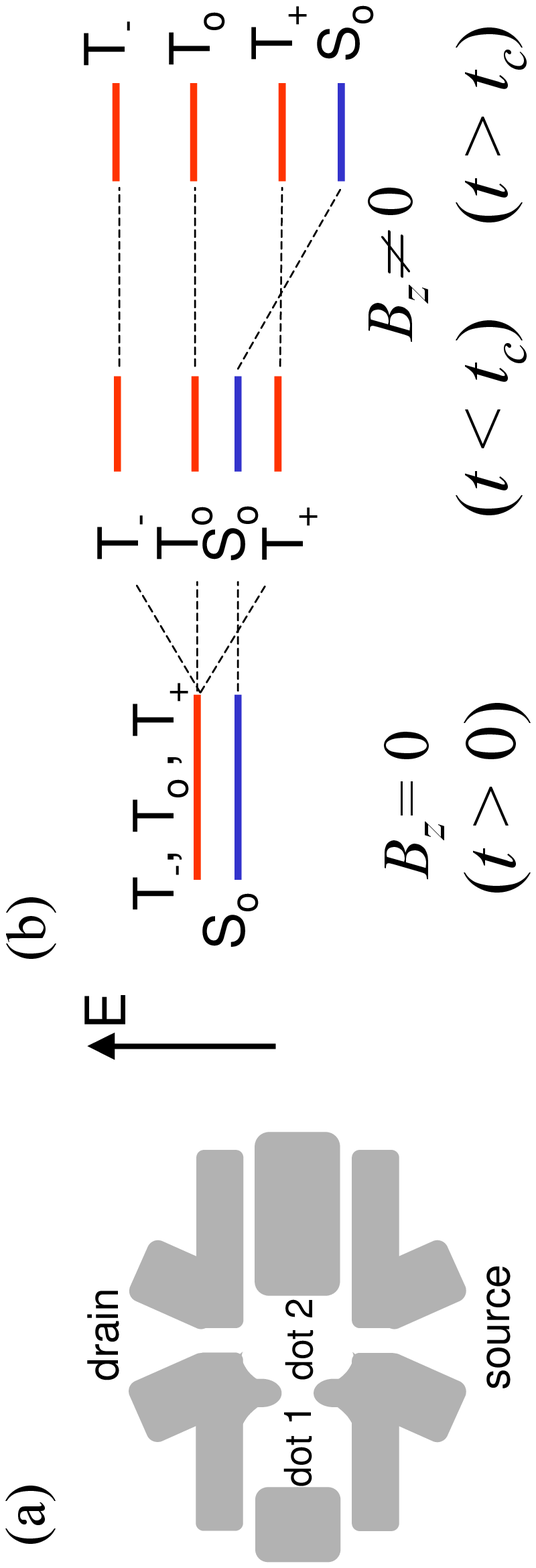}
 \caption{(Color online) (a) Schematic of the double quantum dot structure.
Current passes `only through' dot 2; dots are connected via a
gated quantum point contact with tunneling $t$.  (b) Low energy
level structure for two electrons in a DQD in various regimes.
Notice singlet is the ground state for $t > t_c$, as shown on
right.  } \label{fig1}
\end{figure}

The Hamiltonian of the DQD molecule is $H=H_0 + H_{int}$, where
\begin{eqnarray}
H_0 &=& \sum_{i,\sigma} \epsilon_i c_{i,\sigma}^\dagger
c_{i,\sigma}  -t
\sum_{i\ne j,\sigma} c_{i,\sigma}^\dagger c_{j,\sigma}
\end{eqnarray}
\vspace{-0.3 cm}

\noindent and
\vspace{-0.3 cm}
\begin{eqnarray}
H_{int} &=& U\sum_{i}\hat n_{i,\uparrow}\hat n_{i,\downarrow}+ \frac{1}{2}\left ( V-\frac{J}{2}\right )\sum_{i\ne j}\hat n_i \hat n_j\nonumber \\   
&-& J\sum_{i\ne j}{\bold S}_{i}\cdot {\bold S}_{j}+ \frac{1}{2}\sum_{i,\sigma,\sigma'} \Delta_z^{(i)}
c_{i,\sigma}^\dagger (\sigma_z)_{\sigma\sigma'}c_{i,\sigma'} \, ,  
\end{eqnarray}

\noindent where $c_{i,\sigma}^\dagger$ is the creation operator
for electrons on each dot $i$ with spin $\sigma$
$(\uparrow,\downarrow)$, $t$ describes the interdot orbital
hybridization, which can be tuned in a typical setup by voltages
on the quantum point contact defining the connection between dots.
 In (2), $U$ is the double occupation charging
energy, $V$ is the interdot Coulomb interaction, $J$ gives the interdot exchange interaction (with $J>0$)\cite{Entin-Wohlman}, and finally $\Delta_z^{(i)} =
|g|\mu_B B_z^{(i)}$ is the Zeeman energy splitting for dot $i$ in
a local magnetic field $B_z^{(i)}$. Such local splitting could be
produced for instance by nanoscale magnetic disks of ferromagnetic
material, \cite{Shinjo-Wachowiak} as suggested recently.
\cite{Berciu}. Spin-orbit coupling is also
introduced as in [\onlinecite{Mireles}] but its strength is
typically small ($ t_{SO} < 0.1t $), and does not appreciably
affect our results and conclusions.

We are interested in the effective low-occupation of the DQD, so
that the relevant basis (levels close to the Fermi energy)
consists of a few states for $N=0,1$ and $2$ electrons.

Consider now the case with symmetrical QD's
($\epsilon_1=\epsilon_2=\epsilon$), and homogeneous magnetic
field, $\Delta_z^{(1)}=\Delta_z^{(2)}=\Delta_z$. A non-zero
magnetic field breaks the spin degeneracy in the DQD.\@ The energy
spectrum for $N=1$ is given by $ E^{(1)}_{1,2}=\epsilon -t \mp
\Delta_z /2 $, and $ E^{(1)}_{3,4}=\epsilon + t \mp \Delta_z /2 $,
with (non-normalized) eigenstates $|\upsilon_1
\rangle=|\uparrow,0\rangle + |0,\uparrow\rangle$, $|\upsilon_2
\rangle=|\downarrow,0\rangle + |0,\downarrow\rangle$, $|\upsilon_3
\rangle=|0,\uparrow\rangle - |\uparrow,0\rangle$, and $|\upsilon_4
\rangle=|0,\downarrow\rangle - |\downarrow,0\rangle$,
respectively. For two electrons in the DQD, the magnetic field
breaks the degeneracy of the triplet states
$|T_o\rangle=|\uparrow,\downarrow\rangle +
|\downarrow,\uparrow\rangle  $,
$|T_+\rangle=|\uparrow,\uparrow\rangle$ and
$|T_-\rangle=|\downarrow,\downarrow\rangle$.  The competition
between the Zeeman energy, the exchange  and the interdot hybridization strength
produces a singlet-triplet transition {\em for the ground state},
as shown schematically in Fig.\ 1b. For a fixed (large enough)
magnetic field and $t<t_c$, the ground state is the triplet
$|T_{+}\rangle=|\uparrow,\uparrow\rangle$ with energy
$E^{(2)}_{1}({t<t_c})=2\epsilon +V -\Delta_z - J$, while for
$t>t_c$ the ground state is the singlet configuration
$|S_{o}\rangle=|\uparrow,\downarrow\rangle -
|\downarrow,\uparrow\rangle +\alpha (|\uparrow\downarrow,0\rangle
- |0,\downarrow\uparrow\rangle)$,
%\cite{notasimetria} 
where $\alpha$ is $t$-dependent,
$\alpha=4t/(U-V-J+\sqrt{16t^2+(U-V-J)^2})$.  The singlet has a
field-independent energy $E^{(2)}_{1}({t>t_c})= 2\epsilon
+\frac{1}{2}(U+V+J-\sqrt{16t^2 +(U-V-J)^2})$.  The transition
occurs then at $t=t_c=\frac{1}{2}\sqrt{(2J+\Delta_z)
(U-V+\Delta_z +J)}$.

The conductance per spin in the low bias regime is given by
\cite{Beenakker,Klimeck}
\begin{eqnarray}
G_\sigma(T,V_g)&=&\frac{e^2}{k_B
T}\sum_{n=1}^{N_{max}}\sum_{\alpha,\beta}
\frac{\Gamma_{n\alpha\beta}^{L\sigma}\Gamma_{n\alpha\beta}^{R\sigma}}%
{\Gamma_{n\alpha\beta}^{L\sigma}+\Gamma_{n\alpha\beta}^{R\sigma}}
P_{n,\alpha}^{eq} \nonumber \\
&&\times [1-f_{FD}(E_{n,\alpha}-E_{n-1,\beta}-\mu )] \label{G-eq}
\end{eqnarray}
where $\Gamma_{n\alpha\beta}^{\nu\sigma}=\gamma_{\nu\sigma}%
S_{\alpha\beta}^{\nu\sigma}$ is the tunneling rate involving lead
$\nu$ for an electron with spin $\sigma$. $\,\gamma_{\nu\sigma}$
measures the tunneling amplitude from/to the leads, while
$S_{\alpha\beta}^{\nu\sigma}= |\langle %
n,\alpha|c_{\sigma}^\dagger|n-1,\beta\rangle|^2$ is the spectral
weight whereby the DQD goes from a quantum state $\beta$ with
$n-1$ electrons to a quantum state $\alpha$ with $n$ electrons
(see references for details). \cite{Beenakker,daniela,ramirez}
This conductance expression assumes weak coupling to the leads and
sufficiently high temperatures that one can neglect the Kondo
effect due to correlations with the reservoir electrons.
\cite{Chen}

The only non-zero contributions to the spectral weights involving
the DQD ground state, for the $N:0\rightarrow 1$ transition are
$S_{11}^{\nu,\uparrow}=|\langle
\upsilon_1|c_{\uparrow}^\dagger|0\rangle|^2$, while, for
$N:1\rightarrow 2$ one gets $S_{11}^{\nu,\uparrow}=|\langle
T_+|c_{\uparrow}^\dagger|\upsilon_1\rangle|^2$, for $t<t_c$, and
$S_{11}^{\nu,\downarrow}=|\langle
S_0|c_{\downarrow}^\dagger|\upsilon_1\rangle|^2$, for $t>t_c$.
This $t$-dependence of the spectral weights and the non-trivial
contribution of the first excited states results in spin polarized
transport through the DQD for $N>1$, as we show below.

\begin{figure}[htb]
\begin{center}
%\vspace{-0.2in}
\includegraphics[width=2.6in]{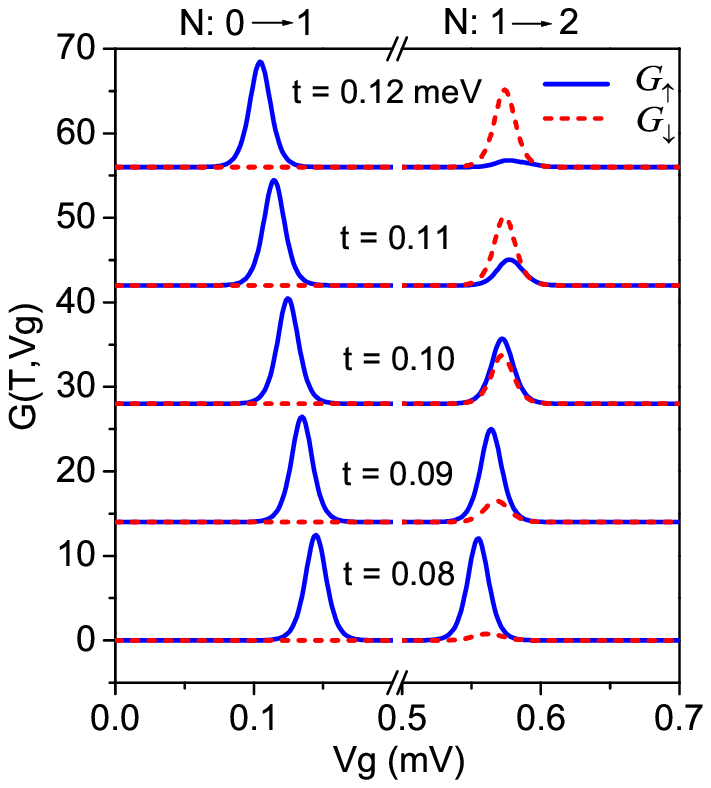}
\vspace{-2em}
\end{center}
 \caption{(Color online) Spin-resolved conductance (in arb. units)
for up (blue solid) and down (red dashed) electrons, as function
of the interdot hopping amplitude $t$ in a magnetic field.  Notice
that for $t>t_c=0.1meV$, the second (higher $V_g$) Coulomb
blockade conductance peak changes to spin-down, while the first
Coulomb blockade peak is always spin-up.  This is the bipolar
filter effect. Here, $t_{SO}=0.1 t$.} \label{fig2}
\end{figure}

% fig 2

Figure 2 shows the conductance through the DQD molecule as
function of the gate voltage $V_g$ for various interdot tunneling
(hybridization) strengths $t$. The Coulomb interdot energy is set
here to $V=0.24meV$, with $\epsilon_1=\epsilon_2=0.25meV$,
$U=1meV$,\cite{charging_energy} $J = 0.01meV$ for the DQD
sizes of interest, \cite{Hu_DasSarma}
$\Delta_z = 0.03 meV$, and symmetrical
tunneling to/from the leads $\gamma_{\nu \sigma}=\Gamma_o =
0.25\mu eV$. At low interdot tunneling, the conductance is
completely spin polarized for the transitions $N:0\rightarrow 1$
and $1 \rightarrow 2$ in the DQD.\@ Since transport in this low
bias regime is determined by the ground state, the spin-up
polarized conductance demonstrates that the Zeeman energy
dominates at these parameters and the conductance accesses the
{\em triplet} ground state, as one would expect. However, as the
interdot tunneling increases, the electrons delocalize, increasing
the hybridization and making the {\em singlet} configuration the
ground state of the DQD.\@ Correspondingly, the second conductance
peak at higher gate voltage, for the transition $N:1 \rightarrow
2$, has a reversed spin character with respect to the first. The
conductance changes from spin {\em up} polarization at weak
coupling ($t <  0.08$) to spin {\em down} polarization at higher
$t$ values ($t\ge 0.12$), with a crossover point a $t_c=0.1meV$,
where both spin conductances are nearly identical. The large
interdot hybridization results in the switch of the spin
polarization for the current through the DQD, resulting in a {\it
bipolar} spin filter with high efficiency (up to $\sim 80 \%$);
all by just adjusting the quantum point contact between dots.

The bipolar spin filter is robust to detuning of the on-site
energies $\Delta\epsilon=\epsilon_2-\epsilon_1$. For example,
while fixing $\epsilon_2$ and varying $\epsilon_1$ via local gates
on each dot, the bipolar function is basically unaffected for
$\Delta_z \ge |\Delta\epsilon| $, except for slight
shifts in the overall position of the Coulomb blockade peaks (not
shown). In contrast, the spin filter effect can be strongly
modified by large asymmetries in the {\em local} Zeeman splitting.
This flexibility might be useful if one controls the local
effective field in the system, via magnetic disks placed in close
proximity to the dots, or local gating to affect the
individual-dot $g$-factors, for example.

We have also explored the DQD in the non-linear regime of
transport. The spin dependent current is calculated by
generalizing Eq.\ (3) of Ref.\ \onlinecite{daniela}. Figure 3
shows the spin polarization current map
($I_{\uparrow}-I_{\downarrow}$) in the $V_{sd}-V_g$ plane, where
$V_{sd}$ is the source-drain bias. Panel (a) shows a net electron
spin current through the DQD for relatively large tunneling
strength, $t=0.15meV$, whereas lower panel (b) depicts the case of
weak tunneling, $t=0.01meV$.  In both cases the first electron
traverses the molecule with spin up (red region, indicated by an
up arrow), given the spin-polarized nature of the ground state,
for a range of $V_{sd}$ until the first excited state enters the
conducting window and the spin polarization is reduced (blue
region).  At a given $V_{sd}$ value, increasing $V_g$ results then
in a drop in the spin polarization, with possible alternations in
value (as in panel (a)).  A larger interdot coupling $t$ results
in a larger $N=1$ Coulomb diamond, as the `bonding-antibonding'
gap increases. The width of spin-up polarized current `bands,' for
the $N:0\rightarrow 1$ transition (red regions), is given by
$\Delta_z$, as we would expect. Also, the separation
between these bands for either fixed $V_g$ or $V_{sd}$ is simply
$2t$. For the spin-down polarized current, this width is
approximately $\Delta_z$, as one can verify from the
energy expressions above.

For small $t$ values, as in panel (b), successive bands of
polarized current are either up or near zero polarization,
indicative of the predominant triplet states at or near the ground
state.  On the other hand, for $t> t_c$, as in panel (a),
increasing $V_g$ results eventually in a band of down-polarized
current (yellow region, indicated by a down arrow), which extends
over a large region in the $V_{sd}-V_g$ plane.  In this regime
the ground state is given by the singlet, the result of the
increased hybridization of the DQD molecular state.

%fig 3
\begin{figure}[htb]
\begin{center}
\vspace{-0.2 in}
\includegraphics[width=2.8in]{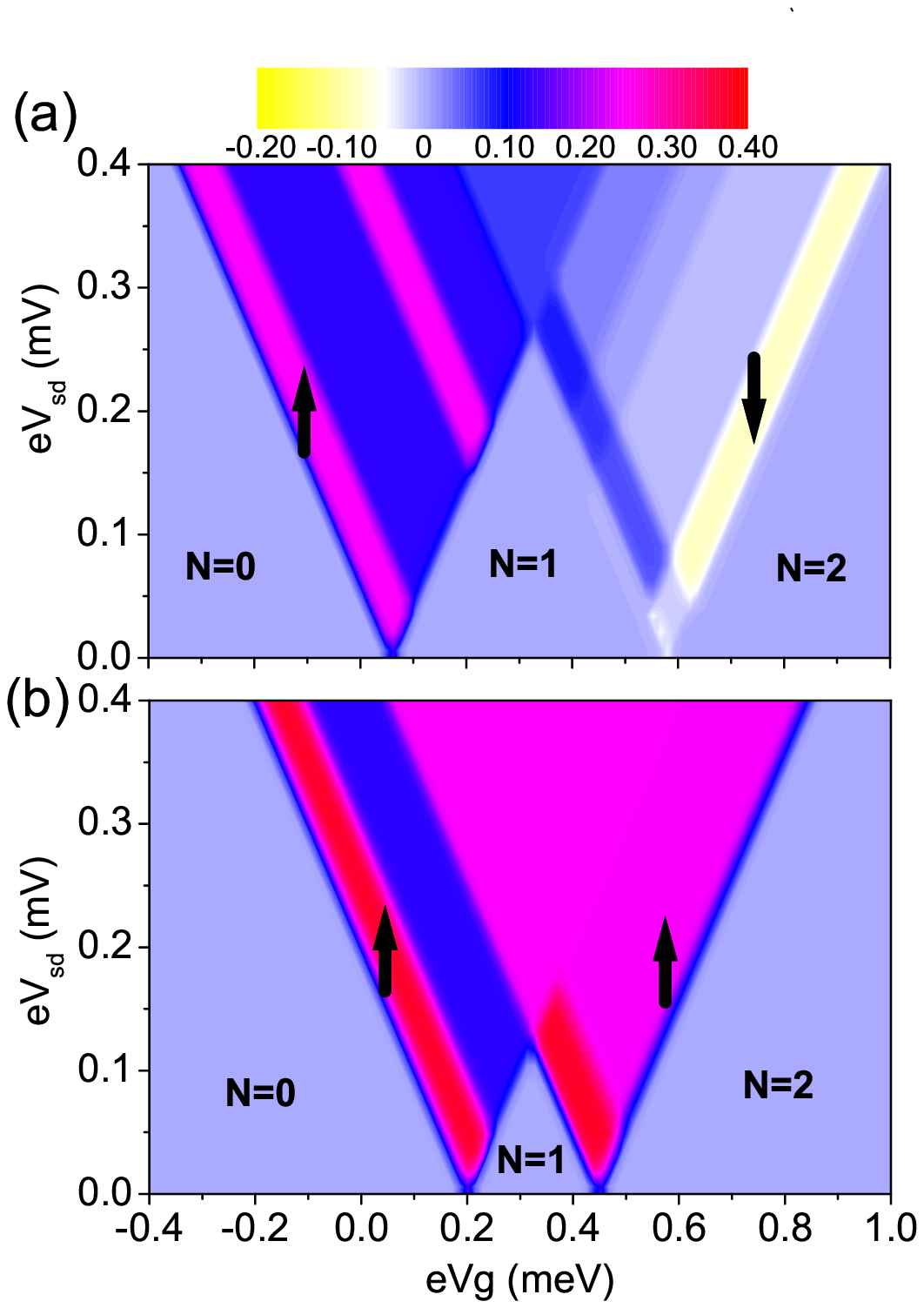}
\vspace{-2em}
\end{center}
\caption{(Color online) Map of the current spin polarization
($I_{\uparrow}-I_{\downarrow}$) through a DQD molecule conducting
through a single dot; Zeeman splitting $\Delta_z =
0.08meV$. (a) Results for $t = 0.15meV > t_c$. Notice both bands of up (red, up arrow) and down (yellow, down arrow) spin. (b) For
small interdot coupling, $t=0.01meV$, only up spins dominate the
current at all bias and gate conditions.  Other parameters as in
Fig.\ 2.} %\vspace{-2em}
\end{figure}

Although our calculations here are for the first few electrons in
the DQD molecule, our discussion should be valid for any
relatively isolated manifold of the molecule level structure.  In
other words, low-lying states have zero spin (typical of any
closed shell), and only states near the Fermi level would
mix/hybridize, reproducing the regime we discuss here.

This work was supported in part by CONACyT-Mexico projects J40521F
and 43673, DGAPA-UNAM project 114403, and the 21$^{st}$ Century
Indiana Fund.

%\newpage 
%{\Large\bf References}


\begin{thebibliography}{99}
\bibitem{Wolf}
S. A. Wolf {\it et al.}, Science {\bf 294}, 1488 (2001).

% S. A. Wolf, D. D. Awschalom, R. A. Buhrman, J. M. Daughton, S. von
% Molnar, M. L. Roukes, A. Y. Chtchelkanova, and D. M. Treger,
% Science {\bf 294}, 1488 (2001).

% Spintronics: A Spin-Based Electronics Vision for the Future

\bibitem{OhnoDMS}
Y. Ohno {\it et al.}, Nature (London) {\bf 402}, 790 (1999);
Fiederling et al., Nature (London) {\bf 402}, 787 (1999).

% Y. Ohno, D.K. Young, B. Beschoten, F. Matsukura, H. Ohno and D.D.
% Awschalom, Nature (London) {\bf 402}, 790 (1999), and Fiederling et al.,
% Nature 402, 787 (1999).

\bibitem{ciorga}
M. Ciorga {\it et al.}, Appl. Phys. Lett. {\bf 80}, 2177 (2002).

% M. Ciorga, M. Pioro-Ladriere, P. Zawadski, P. Hawrylak, and A.S.
% Sachrajda, Appl. Phys. Lett. {\bf 80}, 2177 (2002).


\bibitem{Recher}
P. Recher {\it et al.}, Phys. Rev. Lett. {\bf 85}, 1962 (2000).


% P. Recher, E. V. Sukhorukov, and D. Loss, Phys. Rev.
% Lett. {\bf 85}, 1962 (2000).

\bibitem{Hanson}
R. Hanson {\it et al.}, Phys. Rev. B Rapid Comm. {\bf 70}, 24130 (2004).

% R. Hanson, L.M.K. Vandersypen, L.H. Willems van Beveren, J.M.
% Elzerman, I.T. Vink, and L.P. Kouwenhoven, Phys. Rev. B Rapid
% Comm. {\bf 70}, 24130 (2004)

\bibitem{Folk-Science} J. A. Folk {\it et al.}, Science {\bf 299}, 679 (2003).

%J. A. Folk, R. M. Potok, C. M. Marcus, and V. Umansky, Science {\bf 299}, % 679 (2003).
% A Gate-Controlled Bidirectional Spin Filter Using Quantum Coherence

\bibitem{Potok1} R. M. Potok {\it et al.}, Phys. Rev. Lett. {\bf 89}, 266602 (2002).

%R. M. Potok, J. A. Folk, C. M. Marcus, and V. Umansky,
% Phys. Rev. Lett. {\bf 89}, 266602 (2002)


\bibitem{Potok2} R. M. Potok {\it et al.}, Phys. Rev. Lett. {\bf 91}, 016802 (2003).

%R. M. Potok, J. A. Folk, C. M. Marcus, V. Umansky, M. Hanson,
% and A. C. Gossard, Phys. Rev. Lett. {\bf 91}, 016802 (2003)

\bibitem{Ono} K. Ono {\it et al.}, Science {\bf 297}, 1313 (2002).

% K. Ono, D.G. Austing, Y. Tokura, and S. Tarucha, Science {\bf 297}, 1313 % (2002).

\bibitem{Johnson} A. C. Johnson {\it et al.},  Phys. Rev. B {\bf 72},165308 (2005). % out yet? YES

% A. C. Johnson, J. R. Petta, C. M. Marcus, M. P. Hanson, and
% A.C. Gossard, cond-mat/0410679 (2004). % out yet? YES

\bibitem{Pioro} M. Pioro-Ladriere {\it et al.}, Phys. Rev. Lett. {\bf 91},
026803 (2003).

\bibitem{Entin-Wohlman} O. Entin- Wohlman {\it et al.}, Phys. Rev. B {\bf 64}, 085332 (2001).


% M. Pioro-Ladriere, M. Ciorga, P. Zawadzki, M. Korkusinski, P. Hawrylak,
% \biand A. S. Sachrajda, Phys. Rev. Lett. {\bf 91}, 026803 (2003).

\bibitem{Shinjo-Wachowiak} T. Shinjo {\it et al.}, Science {\bf 289},
930 (2000); A. Wachowiack {\it et al.}, Science {\bf 298}, 577 (2002).


% T. Shinjo, T. Okuno, R. Hassdorf, K. Shigeto,
% and T. Ono, Science {\bf 289}, 930 (2000); A. Wachowiack, J.
% Wiebe, M. Bode, O. Pietzsch, M. Morgenstern, and R. Wiesendanger,
% Science {\bf 298}, 577 (2002).


\bibitem{Berciu} M. Berciu and B. Jank\'o, Phys. Rev. Lett. {\bf 90}, 246804 (2003).

\bibitem{Mireles}
F. Mireles and G. Kirczenow, Phys. Rev B {\bf 64}, 024426 (2001).

%\bibitem{notasimetria}
%With the definitions $|\uparrow \downarrow, 0\rangle = c^{\dagger}_{1,\uparrow}c^{\dagger}_{1,\downarrow}|0\rangle$, and $|0,\uparrow \downarrow\rangle = c^{\dagger}_{2,\uparrow}c^{\dagger}_{2,\downarrow}|0\rangle$

\bibitem{Beenakker}
C.W.J. Beenakker, Phys. Rev. B {\bf 44}, 1646 (1991).

\bibitem{Klimeck}
G. Klimeck {\it et al.}, Phys. Rev. B {\bf 50}, 2316(1994).

% G. Klimeck, G. Chen, and S. Datta, Phys. Rev. B {\bf 50}, 2316(1994).


\bibitem{daniela}
D. Pfannkuche and S. E. Ulloa, Phys. Rev Lett. {\bf 74}, 1194
(1995).

\bibitem{ramirez}
F. Ramirez {\it et al.}, Phys. Rev. B {\bf 59}, 5717 (1999).

% F. Ramirez , E. Cota and S. E. Ulloa, Phys. Rev. B {\bf 59}, 5717 (1999).


\bibitem{Chen} J.C. Chen {\it et al.}, Phys Rev. Lett. {\bf 92}, 176801 (2004).

% J.C. Chen, A.M. Chang, and M.R. Melloch, Phys Rev. Lett.
% {\bf 92},176801 (2004).


\bibitem{charging_energy} Charging energy $U=1meV$, corresponds to
GaAs quantum dots of size $\sim 110$nm. Such value results in
single level spacing $\Delta E \sim 1.7meV$, a value that lies
within the range of recent experiments in small quantum dots, [see
e.g. [5], and Fujisawa {\em et al}., Nature {\bf 419}, 278
(2002)]. Hence $\Delta E >> t$, and neglecting level mixing is an
apropriate approximation.

\bibitem{Hu_DasSarma} X. Hu and S. Das Sarma, \pra {\bf 61},
062301 (2000).

\end{thebibliography}
\end{document}